\documentclass[12pt]{iopart}
\usepackage{amssymb}
\usepackage{graphicx}
\usepackage{dcolumn}
\usepackage{bm}
\usepackage{subfigure}
\usepackage{xcolor}

\begin{document}

\title{The best approximation of the given qubit state with the limited pure-state set}

\author{Li-qiang Zhang$^1$, Deng-hui Yu$^1$, and Chang-shui Yu*$^{1,2}$}

\address{$^1$School of Physics, Dalian University of Technology, Dalian 116024, China}
\address{$^{2}$DUT-BSU Joint Institute, Dalian University of Technology, Dalian, 116024, China}
\ead{ycs@dlut.edu.cn}
\vspace{12pt}
\begin{indented}
\item[]
\end{indented}

\begin{abstract}
The preparation of quantum states lies at the foundation in the 
quantum information processing. The convex mixing of some
existing quantum states is one of the effective candidate. In this paper, we mainly study how a target
quantum state can be optimally prepared by not more than three given pure states.
The analytic optimal distance based on the fidelity is found. We also show
that the preparation with more than four states can be essentially converted
to the case with not more than four states, which can be similarly solved as
the case with three states. The validity is illustrated by the comparison of
our analytical and numerical results.
\end{abstract}
\submitto{\jpa}
\maketitle

%
%
%
%
%

\section{Introduction}

Resource theories are an indispensable part of quantum information theory,
because they can identify resource states that are potentially useful for
quantum information processing. The quantification of quantum resource is
the heart of the resource theory, which has attracted the more and more
attention and developed rapidly \cite{R1,R2,R3,R4,R5,R6}. One important
method to quantifying quantum resource is to measure the closest distance
between the resource state and the free states. For example, quantum
entanglement can be regarded as the distance between the considered quantum state
and all potential convex mixings of the separable states \cite{E01,E02,F1,
E1, E2,E3}. Quantum discord can be quantified by the minimal distance
between the given state and the quantum-classical (or classical-quantum or
classical-classical) states \cite{D4,D5,D6,D7,D8,D9,D10,D11,D12,D13,D14}.
Analogously, quantum coherence can be measured by the closest distance
between the given state and the convex mixings of some given orthonormal
basis \cite{C10,C1,C11,C2,C3,C4,C6,C7,C8,C12,C13}.

Recently, the generalized problem, optimally approximating a desired and
unavailable quantum channel $\Phi $ (quantum state $\rho $) by the convex
mixing of a given set of other channels $\{\Psi _{i}\}$ (quantum states $%
\{\chi _{i}\}$) was addressed in Ref. \cite{CC1,CC2,CC3,CC4}. The convex
approximation problem is to seek for the least distinguishable channel
(quantum state) from $\Phi $($\rho $) among the convex set $%
\sum_{i}p_{i}\Psi _{i}$ ($\sum_{i}p_{i}\chi _{i}$). In Ref. \cite{CC2}, the
author studied the problem based on the $B_{3}$-distance. Namely, the
disposable quantum states come from six eigenstates of Pauli matrices and
the distance metric is taken as the trace norm. In addition, the $B_{2}$%
-distance between any target state and the eigenstates of two of three Pauli
matrices was considered in Ref. \cite{CC3,CC4}.

In this paper, we extend the available quantum states from the eigenstates
of Pauli matrices to any qubit state \cite{T3,T2,T1,T4,T5}. Here we
quantify the distance between two qubit states based on the fidelity. We
analytically investigate the optimal approximation with not more than three
given pure states and find the closest distance, but the similar approach
can be used to the case with four pure states, however, the procedure could
be too tedious to present. In addition, we show that the case with more than
four pure states can be converted to the cases with not more than four states. Our
analytic results are verified by the comparison with the numerical
computations. The remaining of this paper is organized as follows. In Sec.
II, we present the analytic expression of the optimal approximation problem.
In Sec. III, we consider several randomly generated examples to test our
analytic results. The conclusion is obtained in Sec. IV.

\section{The analytic closest distance}

In order to measure the distance of two states, one will have to select a
good metric to unambiguously discriminate two states. Here we employ the
fidelity to assistantly quantify the state distance, so the distance between
any two given quantum states $\rho $ and $\sigma $ can be defined as
\begin{center}
\begin{equation}
D\left( \rho ,\sigma \right) =1-F\left( \rho ,\sigma \right) ,
\end{equation}
\end{center}
where $F\left( \rho ,\sigma \right) $ =Tr$\sqrt{\rho ^{1/2}\sigma \rho ^{1/2}%
}$ is the Fidelity. It is obvious that $D\left( \rho ,\sigma \right) $
inheriting the good properties of Fidelity is (joint) convex and contractive
under trace preserving quantum operation. $D\left( \rho ,\sigma \right) =0$
for $\rho =\sigma $ and $D\left( \rho ,\sigma \right) =1$ for $\rho \perp
\sigma $. Thus the main problem addressed in the paper can be summarized as $%
\min_{\vec{p}}D\left( \rho ,\chi _{1,2,\cdots ,N}\left( \vec{p}\right)
\right) $ with $\chi _{1,2,\cdots ,N}\left( \vec{p}\right)
=\sum\limits_{i=1}^{N}p_{i}\left\vert \varphi _{i}\right\rangle \left\langle
\varphi _{i}\right\vert $ and $\chi _{1,2,\cdots ,N}\left( \vec{p}\right)
=\sum\limits_{i=1}^{N}p_{i}\left\vert \varphi _{i}\right\rangle \left\langle
\varphi _{i}\right\vert $ such that $\sum\limits_{i=1}^{N}p_{i}=1$ for $%
p_{i}\geqslant 0$. Obviously, this main problem can be converted to another
equivalent one as follows.

\noindent Given an objective state of qubit $\rho $ and a pure-state set $%
S:=\left\{ \left\vert \varphi _{i}\right\rangle ,i=1,2,\cdots ,N\right\} $,
the optimal approximation problem reads
\begin{equation}
(\mathbf{P}):\left\{
\begin{array}{ccc}
& \min_{\vec{p}}-F^{2}\left( \rho ,\chi _{1,2,\cdots ,N}\left( \vec{p}%
\right) \right) , &  \\
\mbox{subject to}: & \sum\limits_{i=1}^{N}p_{i}=1, & p_{i}\geqslant 0%
\end{array}%
\right. ,  \label{D1}
\end{equation}%
where $\chi _{1,2,\cdots ,K}\left( \vec{p}\right)
=\sum\limits_{i=1}^{K}p_{i}\left\vert \varphi _{i}\right\rangle \left\langle
\varphi _{i}\right\vert $ represents the optimized state convexly mixed by
the $1$st, $2$st, $\cdots ,$ $K$th states ($N\geqslant K$) subject to the
set $S$.

To proceed, we'd like to consider the question in the Bloch representation.
Let $\mathbf{r}_{o}$ denote the Bloch vector of $\rho $ with its elements $%
r_{o\alpha }=$ Tr$(\rho \sigma _{\alpha })$ and $\mathbf{r}_{k}$ ($%
r_{k\alpha }$ is its element) denote the Bloch vector of the $k$th state in $%
S$, where $\sigma _{\alpha }$ are Pauli matrices with $\alpha =x,y,z$. Thus
the squared fidelity between the qubit states $\rho $ and $\sigma =\chi
_{1,2,\cdots ,K}\left( \vec{p}\right) $ is given by
\begin{equation}
F^{2}\left( \rho ,\chi \right) =\frac{1}{2}+\sum_{i=1}^{N}p_{i}\frac{\mathbf{%
r}_{o}^{T}\mathbf{r}_{i}}{2}+\sqrt{\frac{\mathfrak{m}}{2}}\sqrt{s},
\label{fsq}
\end{equation}%
where $s=\mathbf{p}^{T}Y\mathbf{p}$ with $\mathbf{p}=\left(
p_{1},p_{2},p_{3},\cdots ,p_{K}\right) ^{T}$, and $\mathfrak{m}=1-$Tr$\rho
^{2}$ representing the mixedness of $\rho $. In particular, in Eq. (\ref{fsq}%
), $Y_{ij}=1-\mathbf{r}_{i}^{T}\mathbf{r}_{j}$ denotes the distance between
two the qubit states $\mathbf{r}_{i}$ and $\mathbf{r}_{j}$. In this sense, the
difference of the two distances from the two states to the objective state
is $M_{ij}=\mathbf{r}_{o}^{T}\left( \mathbf{r}_{i}-\mathbf{r}_{j}\right) $.
With these notations, the problem (P) can be solved as follows.

\textbf{Theorem 1}.- If there are two pure states ($N=2$) in the set $S$,
one can construct an qubit state $\chi _{1,2}\left( \vec{p}\right)$ with the
optimal distance given by%
\begin{equation}
D(\rho ,\chi _{1,2}\left( \vec{p}\right) )=1-F(\rho ,\chi _{1,2}\left( \vec{p%
}\right) ),  \label{J1}
\end{equation}%
where the optimal fidelity $F(\rho ,\chi _{1,2}\left( \vec{p}\right) )$ is
\begin{equation}
F^{2}(\rho ,\chi _{1,2}\left( \vec{p}\right) )=\frac{2M_{+}+\sqrt{4\mathfrak{%
m}Y_{12}+M_{12}^{2}}}{4}  \label{J1a}
\end{equation}%
with $M_{+}=1+\frac{\mathbf{r}_{o}^{T}\left( \mathbf{r}_{1}+\mathbf{r}%
_{2}\right) }{2}$. In particular, the optimal weight $\vec{p}$ which
achieves the above optimal distance reads
\begin{equation}
p_{1}=\frac{1}{2}[1+\frac{M_{12}}{\sqrt{4\mathfrak{m}Y_{12}+M_{12}^{2}}}].
\end{equation}

\textbf{Proof}. For $N=2$, the squared fidelity in Eq. (\ref{fsq}) is given
by
\begin{eqnarray}
F^{2}(\rho ,\chi _{1,2}\left( \vec{p}\right) )&=\sum_{i=1}^{2}p_{i}\frac{1+%
\mathbf{r}_{o}^{T}\mathbf{r}_{i}}{2}+\sqrt{\mathfrak{m}Y_{12}}\sqrt{p_{1}p_{2}}     \cr
&=\frac{M_{12}p_{1}}{2}+\sqrt{\mathfrak{m}Y_{12}}\sqrt{p_{1}-p_{1}^{2}}+\frac{%
1+\mathbf{r}_{o}^{T}\mathbf{r}_{2}}{2}.  \label{D2}
\end{eqnarray}
The maximum value of $F^{2}$ can be solved by the Lagrange multiplier
approach. The derivative of $F^{2}(\rho ,\chi _{1,2}\left( \vec{p}\right) )$
concerning the parameters $p_{1}$ is given by%
\begin{equation}
\frac{\partial F^{2}}{\partial p_{1}}=\frac{1}{2}\left[ M_{12}+\frac{1-2p_{1}%
}{\sqrt{p_{1}(1-p_{1})}}\sqrt{\mathfrak{m}Y_{12}}\right] .
\end{equation}%
Solving $\frac{\partial F^{2}}{\partial p_{1}}=0$ will arrive at a valid $%
p_{1}$ as$\ $
\begin{eqnarray}
p_{1} =\frac{1}{2}[1+\frac{M_{12}}{\sqrt{4\mathfrak{m}Y_{12}+M_{12}^{2}}}],p_{2} =1-p_{1}.
\label{DD2}
\end{eqnarray}%
Inserting $p_{i}$ into $\rho _{0}=\chi _{1,2}\left( \vec{p}\right) $, one
will find
\begin{equation}
F^{2}(\rho ,\rho _{0})=\frac{1}{4}[2M_{+}+\sqrt{4\mathfrak{m}%
Y_{12}+M_{12}^{2}}]
\end{equation}%
with $M_{+}=1+\frac{\mathbf{r}_{o}^{T}\left( \mathbf{r}_{1}+\mathbf{r}%
_{2}\right) }{2}$. The proof is completed.\hfill $\blacksquare $

\textbf{Corollary}.- If the two qubit states$\ \left\vert \varphi
_{1}\right\rangle $ and $\left\vert \varphi _{2}\right\rangle $ in $S$ are
orthonormal, the optimal distance $D(\rho ,\chi _{1,2}\left( \vec{p}\right)
) $ will be reduced to%
\begin{equation}
D(\rho ,\chi _{1,2}\left( \vec{p}\right) )=1-\left[ \frac{1+\sqrt{2\mathfrak{%
m}+(\mathbf{r}_{o}^{T}\mathbf{r}_{i})^{2}}}{2}\right] ^{1/2}  \label{chuizhi}
\end{equation}%
with $i=1,2$. The optimal weight is given by%
\begin{equation}
p_{i}=\frac{1}{2}[1+\frac{\mathbf{r}_{o}^{T}\mathbf{r}_{i}}{\sqrt{2\mathfrak{%
m}+(\mathbf{r}_{o}^{T}\mathbf{r}_{i})^{2}}}].  \label{chuizhip}
\end{equation}%
In particular, $\mathbf{r}_{o}^{T}\mathbf{r}_{i}=r_{o\alpha }r_{i\alpha }$,
if both $\left\vert \varphi _{1}\right\rangle $ and $\left\vert \varphi
_{2}\right\rangle $ are the eigenstates of one Pauli matrix $\sigma _{\alpha
}$ with $\alpha =x,y,z$.

\textbf{Proof}. The mutually orthogonal $\left\vert \varphi
_{1}\right\rangle $ and $\left\vert \varphi _{2}\right\rangle $ means that $%
\mathbf{r}_{1}^{T}\mathbf{r}_{2}=-1$ and $Y_{12}=1-\mathbf{r}_{1}^{T}\mathbf{%
r}_{2}=2$, so $M_{+}=1$, $M_{12}=2M_{\bot }=2\mathbf{r}_{o}^{T}\mathbf{r}%
_{1} $. Substituting these parameters into Theorem 1, one will obtain Eqs. (%
\ref{chuizhi}) and (\ref{chuizhip}).\hfill $\blacksquare $

The above corollary actually gives an alternative quantifier for quantum
coherence of the objective qubit state $\rho $, in which two orthonormal
pure state basis vectors are defined in the set $S$. So Eq. (\ref{chuizhi})
can be used as an analytical solution for the coherence of qubit states,
which coincides with the result in \cite{jiexi}.

\textbf{Theorem 2}.- If there are three pure states ($N=3$) in the set $S$,
one can construct a pseudo-state $\chi _{1,2,3}\left( \vec{p}\right)$ with
the pseudo-probabilities as
\begin{eqnarray}
\fl \tilde{p}_{1} =\frac{1}{4Y_{13}Y_{23}-Y_{123}^{2}}\left[ Y_{23}\left(
Y_{123}+2Y_{13}\right) \right.  -\left( Y_{123}M_{32}+2Y_{23}M_{31}\right) \sqrt{\frac{Y_{12}Y_{13}Y_{23}}{%
\kappa }}],  \label{P1} \\
\fl \tilde{p}_{2} =\frac{1}{4Y_{13}Y_{23}-Y_{123}^{2}}\left[ Y_{13}\left(
Y_{123}+2Y_{23}\right) \right. -\left( Y_{123}M_{31}+2Y_{13}M_{32}\right) \sqrt{\frac{Y_{12}Y_{13}Y_{23}}{%
\kappa }}],  \label{P2} \\
\fl \tilde{p}_{3} =1-\tilde{p}_{1}-\tilde{p}_{2,}  \label{P3}
\end{eqnarray}
where
\begin{equation}
\kappa
=M_{31}^{2}Y_{23}+M_{32}^{2}Y_{13}+M_{31}M_{32}Y_{123}+(4Y_{13}Y_{23}-Y_{123}^{2})%
\mathfrak{m},  \label{kk}
\end{equation}%
with $Y_{123}=Y_{12}-Y_{13}-Y_{23}$.

If $\tilde{p}_{i}\geqslant 0$ holds for $i=1,2,3$, $\chi _{1,2,3}\left( \vec{%
p}\right) $ will be our expected optimal qubit state with the optimal weights $%
p_{i}=\tilde{p}_{i}$, and the optimal distance as%
\begin{equation}
D(\rho ,\chi _{1,2,3}\left( \vec{p}\right) )=1-F(\rho ,\chi _{1,2,3}\left(
\vec{p}\right) ),  \label{J22a1}
\end{equation}%
otherwise, the optimal qubit state and distance are given by%
\begin{equation}
D(\rho ,\chi _{1,2,3}\left( \vec{p}\right) )=\min_{i<j}D(\rho ,\chi
_{i,j}\left( \vec{p}\right) ),i,j=1,2,3,  \label{J22a}
\end{equation}%
which as well as the corresponding weights can be solved by Theorem 1.

\textbf{Proof. }For $N=3$, one can substitute $\chi _{1,2,3}\left( \vec{p}%
\right) $ into $-F^{2}(\rho ,\chi _{1,2,3}\left( \vec{p}\right) )$ and
establish the corresponding Lagrangian function as%
\begin{eqnarray}
L(p_{i},\lambda ,\lambda _{i}) =-\frac{1}{2}-\sum_{i=1}^{3}\frac{\mathbf{r}%
_{o}^{T}\mathbf{r}_{i}}{2}p_{i}-\sqrt{\frac{\mathfrak{m}}{2}}\sqrt{s}  
-\sum_{i=1}^{3}\lambda _{i}p_{i}+\lambda (\sum_{i=1}^{3}p_{i}-1),
\end{eqnarray}%
where $\lambda $ and $\lambda _{i}$ are the Lagrangian multiplier.

\textit{Case 1}\textbf{:} $\mathfrak{m}=0$. In this case, the given
objective state $\rho $ is a pure state. Thus $F^{2}(\rho ,\chi
_{1,2,3}\left( \vec{p}\right) )=\frac{1}{2}+\sum_{i=1}^{3}\frac{\mathbf{r}%
_{o}^{T}\mathbf{r}_{i}}{2}p_{i}$. It is obvious that $\max_{\vec{p}%
}F^{2}(\rho ,\chi _{1,2,3}\left( \vec{p}\right) )=\frac{1+\mathbf{r}_{o}^{T}%
\mathbf{r}_{\max }}{2}$ with $\mathbf{r}_{\max }=\max_{i}\left\{ \mathbf{r}%
_{i}\right\} $, which is included in Eq. (\ref{J22a}).

\textit{Case 2}\textbf{:} $\mathfrak{m}\neq 0.$
The Karush-Kuhn-Tucker conditions \cite{KKT,HH} are given by%
\begin{eqnarray}
\frac{\partial L}{\partial p_{i}} &=&-\frac{\mathbf{r}_{o}^{T}\mathbf{r}_{i}%
}{2}-\sqrt{\frac{\mathfrak{m}}{2}}\frac{\sum_{j\neq i}p_{j}Y_{ij}}{\sqrt{s}}%
+\lambda -\lambda _{i}=0, \nonumber \\
\lambda _{i} &\geq &0,p_{i}\geq 0,\lambda _{i}p_{i}=0,i=1,2,3.   \label{eq} 
\end{eqnarray}
Now let $p_{i}\neq 0$ for $i=1,2,3$ which corresponds to $\lambda _{i}=0$.
Solving Eq. (\ref{eq}) by $\frac{\partial L}{\partial p_{1}}-\frac{\partial L%
}{\partial p_{3}}=0$ and $\frac{\partial L}{\partial p_{2}}-\frac{\partial L%
}{\partial p_{3}}=0$, we have%
\begin{eqnarray}
\left( 1-2p_{1}\right) Y_{13}+Y_{123}p_{2} &=&\frac{M_{31}}{\sqrt{2\mathfrak{%
m}}}\sqrt{s},  \label{D6} \\
Y_{123}p_{1}+\left( 1-2p_{2}\right) Y_{23} &=&\frac{M_{32}}{\sqrt{2\mathfrak{%
m}}}\sqrt{s}.  \label{D61}
\end{eqnarray}%
Eliminating $\sqrt{s}$ from Eqs. (\ref{D6},\ref{D61}) shows%
\begin{eqnarray}
\fl p_{2}(Y_{123}M_{32}+2Y_{23}M_{31}) 
=(Y_{123}M_{31}+2Y_{13}M_{32})p_{1}+Y_{23}M_{31}-Y_{13}M_{32}.
\label{p123}
\end{eqnarray}

\textit{Case 2.1}: $Y_{123}M_{32}+2Y_{23}M_{31}\neq 0$, $%
Y_{123}M_{31}+2Y_{13}M_{32}\neq 0$. From the case, we have%
\begin{eqnarray}
\fl Y_{123}M_{32}+2Y_{23}M_{31}  
=\mathbf{r}_{0}^{T}\left[ (\mathbf{r}_{3}-\mathbf{r}_{1})(\mathbf{r}_{1}-%
\mathbf{r}_{2})^{T}-(\mathbf{r}_{1}-\mathbf{r}_{2})(\mathbf{r}_{3}-\mathbf{r}%
_{1})^{T}\right] (\mathbf{r}_{3}-\mathbf{r}_{2})  
\neq &0  \label{YY1}
\end{eqnarray}%
and%
\begin{eqnarray}
\fl Y_{123}M_{31}+2Y_{13}M_{32} 
=\mathbf{r}_{0}^{T}\left[ (\mathbf{r}_{3}-\mathbf{r}_{2})(\mathbf{r}_{2}-%
\mathbf{r}_{1})^{T}-(\mathbf{r}_{2}-\mathbf{r}_{1})(\mathbf{r}_{3}-\mathbf{r}%
_{2})^{T}\right] (\mathbf{r}_{3}-\mathbf{r}_{1}) 
\neq 0.  \label{YY2}
\end{eqnarray}%
This means that the three quantum states $\{\left\vert \varphi
_{1}\right\rangle ,\left\vert \varphi _{2}\right\rangle ,\left\vert \varphi
_{3}\right\rangle \}$ are different from each other. From Equation (%
\ref{p123}), we can get
\begin{equation}
p_{2}=a_{1}p_{1}+a_{2},  \label{D7}
\end{equation}%
where $a_{1}=\frac{Y_{123}M_{31}+2Y_{13}M_{32}}{Y_{123}M_{32}+2Y_{23}M_{31}}$
and $a_{2}=\frac{Y_{23}M_{31}-Y_{13}M_{32}}{Y_{123}M_{32}+2Y_{23}M_{31}}$.

\textit{Case 2.1.1}: $M_{31}\neq 0$. Substituting  Eq. (\ref{D7}) into
Eq. (\ref{D6}), it follows that%
\begin{equation}
b_{1}p_{1}^{2}+b_{2}p_{1}+b_{3}=0,  \label{eq2}
\end{equation}%
where%
\begin{eqnarray}
b_{1} &=(Y_{123}a_{1}-2Y_{13})^{2}+\frac{M_{31}^{2}}{\mathfrak{m}}  
\lbrack (Y_{23}a_{1}+Y_{13})(a_{1}+1)-Y_{12}a_{1}]   \nonumber \\
&=\frac{M_{31}^{2}\left( 4Y_{13}Y_{23}-Y_{123}^{2}\right) \kappa }{%
(Y_{123}M_{32}+2Y_{23}M_{31})^{2}\mathfrak{m}},
\end{eqnarray}%
\begin{eqnarray}
\fl b_{2} =2(Y_{123}a_{1}-2Y_{13})(Y_{123}a_{2}+Y_{13}) 
+\frac{M_{31}^{2}}{\mathfrak{m}}[(Y_{23}a_{1}+Y_{13})(a_{2}-1)  
+Y_{23}a_{2}(a_{1}+1)-Y_{12}a_{2}]  \nonumber \\
=\frac{-2M_{31}^{2}Y_{23}\left( Y_{123}+2Y_{13}\right) \kappa }{%
(Y_{123}M_{32}+2Y_{23}M_{31})^{2}\mathfrak{m}},
\end{eqnarray}%
\begin{eqnarray}
b_{3} &=&(Y_{123}a_{2}+Y_{13})^{2}+\frac{M_{31}^{2}}{\mathfrak{m}}%
Y_{23}a_{2}(a_{2}-1)  \nonumber \\
&=&\frac{M_{31}^{2}Y_{23}}{(Y_{123}M_{32}+2Y_{23}M_{31})^{2}\mathfrak{m}}
\lbrack (M_{32}^{2}+4Y_{23}\mathfrak{m})Y_{12}Y_{13}-Y_{23}\kappa ],
\end{eqnarray}%
with%
\begin{equation}
\kappa
=M_{31}^{2}Y_{23}+M_{32}^{2}Y_{13}+M_{31}M_{32}Y_{123}+(4Y_{13}Y_{23}-Y_{123}^{2})%
\mathfrak{m}.
\end{equation}%
If $\kappa =0$, we can get%
\begin{eqnarray}
b_{1} &=&0,b_{2}=0,  \nonumber \\
b_{3} &=&\frac{M_{31}^{2}(M_{32}^{2}+4Y_{23}m)Y_{12}Y_{13}Y_{23}}{%
(Y_{123}M_{32}+2Y_{23}M_{31})^{2}m}\neq 0,
\end{eqnarray}%
which contradicts Eq. (\ref{eq2}). So we have $\kappa \neq 0$. Thus the
roots of Eq. (\ref{eq2}) read%
\begin{eqnarray}
\fl \tilde{p}_{1} =\frac{-b_{2}\pm \sqrt{b_{2}^{2}-4b_{1}b_{3}}}{2b_{1}}
\nonumber \\
\fl =\frac{1}{4Y_{13}Y_{23}-Y_{123}^{2}}[Y_{23}\left( Y_{123}+2Y_{13}\right)
\pm \left\vert Y_{123}M_{32}+2Y_{23}M_{31}\right\vert \sqrt{\frac{%
Y_{12}Y_{13}Y_{23}}{\kappa }}].  \label{D8}
\end{eqnarray}%
The two cases ($\pm $) of $\tilde{p}_{1}$ are finally determined by checking
the validity of Eq. (\ref{D6}) by inserting Eq. (\ref{D7}) and Eq. (\ref{D8}%
). And we can get%
\begin{eqnarray}
\fl \tilde{p}_{1} &=&\frac{1}{4Y_{13}Y_{23}-Y_{123}^{2}}[Y_{23}\left(
Y_{123}+2Y_{13}\right) 
-(Y_{123}M_{32}+2Y_{23}M_{31})\sqrt{\frac{Y_{12}Y_{13}Y_{23}}{\kappa }}].
\label{PP1}
\end{eqnarray}%
At the same time, $\tilde{p}_{2}$ can be given by Eq. (\ref{D7}) as%
\begin{eqnarray}
\fl \tilde{p}_{2} &=&\frac{1}{4Y_{13}Y_{23}-Y_{123}^{2}}[Y_{13}\left(
Y_{123}+2Y_{23}\right) -(Y_{123}M_{31}+2Y_{13}M_{32})\sqrt{\frac{Y_{12}Y_{13}Y_{23}}{\kappa }}].
\label{PP2}
\end{eqnarray}%
and $\tilde{p}_{3}=1-\tilde{p}_{1}-\tilde{p}_{2}$. If all $\tilde{p}_{i}$ $%
\in \lbrack 0,1]$, then the optimal weights $p_{i}=\tilde{p}_{i}$. With the
optimal weights $p_{i}$, one can calculate $\chi _{1,2,3}\left( \vec{p}%
\right) =\sum_{i=1}^{3}p_{i}\left\vert \varphi _{i}\right\rangle
\left\langle \varphi _{i}\right\vert $, and the optimal distance is hence
given by%
\begin{equation}
D(\rho ,\chi _{1,2,3}\left( \vec{p}\right) )=1-F(\rho ,\chi _{1,2,3}\left(
\vec{p}\right) ).  \label{TT1}
\end{equation}%
If not all $\tilde{p}_{i}$ $\in \lbrack 0,1]$, this means the optimal
weights should be on the boundaries which are described by (i) $p_{1}\neq 0$%
, $p_{2}\neq 0,\lambda _{3}\neq 0$; (ii)$\ p_{1}\neq 0,\lambda _{2}\neq
0,p_{3}\neq 0$; (iii)$\ \lambda _{1}\neq 0,p_{2}\neq 0,p_{3}\neq 0$. In
other words, there exists at least one $p_{i}=0$ among the three. So the
left two nonvanishing $p_{i}$ convert our original optimization question to
the question of Theorem 1, and then we choose the smallest one as
\begin{equation}
\min_{i<j}D(\rho ,\chi _{i,j}\left( \vec{p}\right) ),i,j=1,2,3,  \label{TT2}
\end{equation}%
which can be exactly solved by Theorem 1.

\textit{Case 2.1.2}: $M_{32}\neq 0$. Substituting Eq. (\ref{D7}) into
Eq. (\ref{D61}), and using a similar process from Eq. (\ref{eq2}) to Eq. (%
\ref{PP2}), we can get the same conclusion as Eq. (\ref{TT1}) and Eq. (\ref%
{TT2}).

\textit{Case 2.2}: $Y_{123}M_{32}+2Y_{23}M_{31}=0$, $%
Y_{123}M_{31}+2Y_{13}M_{32}=0$. So Eq. (\ref{p123}) can be rewritten as%
\begin{equation}
Y_{23}M_{31}-Y_{13}M_{32}=0
\end{equation}%
where the coefficients of $p_{1}$ and $p_{2}$ are zero. This means that for
the homogeneous equation set \{Eq. (\ref{D6}), Eq. (\ref{D61})\}, the rank
of coefficient matrix is less than the number of unknown parameters $\left\{
p_{1},p_{2}\right\} $. That is to say, there is a parameter $p_{i}\in \left[
0,1\right] $ which is a free parameter. Obviously, this case can be solved
by Eq. (\ref{TT2}).

\textit{Case 2.3}: $Y_{123}M_{32}+2Y_{23}M_{31}=0$, $%
Y_{123}M_{31}+2Y_{13}M_{32}\neq 0$. In this case, we can deduce%
\begin{equation}
M_{32}\neq 0,Y_{23}\neq 0.
\end{equation}%
From Eq. (\ref{p123}), we can get
\begin{eqnarray}
\fl \tilde{p}_{1} &=&\frac{Y_{13}M_{32}-Y_{23}M_{31}}{Y_{123}M_{31}+2Y_{13}M_{32}%
}  
=\frac{Y_{13}M_{32}+Y_{123}M_{32}/2}{2Y_{13}M_{32}-Y_{123}^{2}M_{32}/Y_{23}%
}  
=\frac{Y_{23}\left( Y_{123}+2Y_{13}\right) }{4Y_{13}Y_{23}-Y_{123}^{2}}.
\label{P11}
\end{eqnarray}%
After a similar calculation process from Eq. (\ref{eq2}) to Eq. (\ref{PP1}),
one arrives at%
\begin{eqnarray}
\fl \tilde{p}_{2} &=&\frac{1}{4Y_{13}Y_{23}-Y_{123}^{2}}\left[ Y_{13}\left(
Y_{123}+2Y_{23}\right) \right. -\left( Y_{123}M_{31}+2Y_{13}M_{32}\right) \sqrt{\frac{Y_{12}Y_{13}Y_{23}}{%
\kappa }}].
\end{eqnarray}%
Then we can obtain $\tilde{p}_{3}=1-\tilde{p}_{1}-\tilde{p}_{2}$. One can
find that $\{\tilde{p}_{1},\tilde{p}_{2},\tilde{p}_{3}\}$ is just a special
case of Eqs. (\ref{PP1}) and (\ref{PP2}) in \textit{Case 2.1}. So the
optimal distance can be determined equivalently by Eq. (\ref{TT1}) and Eq. (%
\ref{TT2}).

\textit{Case 2.4}: $Y_{123}M_{32}+2Y_{23}M_{31}\neq 0$, $%
Y_{123}M_{31}+2Y_{13}M_{32}=0$. In this case, we can deduce%
\begin{equation}
M_{31}\neq 0,Y_{13}\neq 0.
\end{equation}%
From Eq. (\ref{p123}), we can get%
\begin{eqnarray}
\fl \tilde{p}_{2} =\frac{Y_{23}M_{31}-Y_{13}M_{32}}{Y_{123}M_{32}+2Y_{23}M_{31}%
}  
=\frac{Y_{23}M_{31}+Y_{123}M_{31}/2}{2Y_{23}M_{31}-Y_{123}^{2}M_{31}/Y_{13}%
}  
=\frac{Y_{13}\left( Y_{123}+2Y_{23}\right) }{4Y_{13}Y_{23}-Y_{123}^{2}},
\label{P22}
\end{eqnarray}%
After a similar calculation process from Eq. (\ref{eq2}) to Eq. (\ref{PP1}),
one arrives at%
\begin{eqnarray}
\fl \tilde{p}_{1} &=&\frac{1}{4Y_{13}Y_{23}-Y_{123}^{2}}\left[ Y_{23}\left(
Y_{123}+2Y_{13}\right) \right. 
-\left( Y_{123}M_{32}+2Y_{23}M_{31}\right) \sqrt{\frac{Y_{12}Y_{13}Y_{23}}{%
\kappa }}].
\end{eqnarray}%
Then we can obtain $\tilde{p}_{3}=1-\tilde{p}_{1}-\tilde{p}_{2}$. It is
obvious that $\{\tilde{p}_{1},\tilde{p}_{2},\tilde{p}_{3}\}$ is also covered
by Eqs. (\ref{PP1}) and (\ref{PP2}) in \textit{Case 2.1}. To sum up, one can
find that the formulas given in our theorem actually include all the
discussed special cases. The proof is completed.\hfill $\blacksquare $

Theorem 2 implies some kind of degradation relation. We tend to use the least
number of quantum states to achieve the best approximation. Let's first
consider the case $p_{i}\neq 0$ for $i=1,2,3$. The validity of the pseudo-probabilities of the extreme point in the global range must be guaranteed. If the
pseudo-probabilities are invalid, the problem of $N=3$ will be simplified to
$N=2$, which means that we must use theorem 1 three times to find the
minimum value. The process of finding the minimum value is to satisfy the
condition $\lambda _{i}\geq 0$ in Eq. (\ref{eq}). This degradation relation
ensures that Theorem 2 for $N=3$ is automatically simplified to Theorem 1.
Next, we will focus on the case of $N=4$. To do so, we will first address
the special case of full-rank $4\times K$ matrix $A$, which is defined as $%
A_{ij}=\left\langle \varphi _{j}\right\vert \sigma _{i}\left\vert \varphi
_{j}\right\rangle $ with $j=1,2,3,\cdots ,K$ and $i=1,2,3,4$ corresponding
to $x,y,z$, and $\sigma _{4}=\left(
\begin{array}{cc}
1 & 0 \\
0 & 1%
\end{array}%
\right) $.

\textbf{Theorem 3}.-If there are four pure states ($N=4$) in the set $S$
with the rank-$4$ matrix $A$, we can give the pseudo-probabilities
\begin{equation}
\mathbf{\tilde{p}=}A^{-1}\mathbf{\tilde{r}}
\end{equation}%
where $\mathbf{\tilde{r}=[r};1]$ is a four-dimensional vector. If $\tilde{p}%
_{i}\in \lbrack 0,1]$ for all $i$, the objective qubit state $\rho $ can be
directly written as the convex sum of the four pure states with the optimal
weights $\mathbf{p=\tilde{p}}$.

\textbf{Proof}. Assuming that the target quantum state $\rho $ can be
represented exactly by the linear sum of the four pure states given in $S$,
then we must find the pseudo probability $\tilde{p}_{i}$ such that%
\begin{equation}
\rho =\sum_{i=1}^{4}\tilde{p}_{i}\left\vert \psi _{i}\right\rangle
\left\langle \psi _{i}\right\vert ,  \label{zhijie}
\end{equation}%
with $\sum\limits_{i}\tilde{p}_{i}=1$. Note that $\tilde{p}_{i}$ could be
negative. Thus in the Bloch representation, the above Eq. (\ref{zhijie}) can
be rewritten as
\begin{equation}
A\mathbf{\tilde{p}=\tilde{r}}  \label{pseu}
\end{equation}%
with $\mathbf{\tilde{r}=[r};1\mathbf{]}$ and $\mathbf{\tilde{p}}=\left(
\tilde{p}_{1},\tilde{p}_{2},\tilde{p}_{3},\tilde{p}_{4}\right) ^{T}$ . Since
$A$ is of full rank, Eq. (\ref{pseu}) has a single solution as $\mathbf{%
\tilde{p}=}A^{-1}\mathbf{\tilde{r}}$. It is obvious that if $\tilde{p}%
_{i}\in \lbrack 0,1]$ for all $i$, one will directly obtain the optimal
solution $\mathbf{p}$ of our scheme is $\mathbf{p=\tilde{p}=}A^{-1}\mathbf{%
\tilde{r}}$. The proof is completed. \hfill $\blacksquare $

Theorem 3 actually presents quite limited conclusion for the the problem (P)
with ($N=4$). In fact, the approach in Theorem 2 can be similarly employed
for the case of $N=4$. Obviously, it is so complicated that the analytical and general results like those in Theorem 2 are hard to give. 
So we will only consider some particular pure-state set.
Suppose the set $S$ is made up of the eigenstates of three Pauli matrices. Let
$\left\vert \sigma _{z}\right\rangle _{s}=\left\{ \left\vert 0\right\rangle
,\left\vert 1\right\rangle \right\} $,$\ \left\vert \sigma _{x}\right\rangle
_{s}=\left\{ \frac{1}{\sqrt{2}}(\left\vert 0\right\rangle \pm \left\vert
1\right\rangle )\right\} $ and $\left\vert \sigma _{y}\right\rangle
_{s}=\left\{ \frac{1}{\sqrt{2}}(\left\vert 0\right\rangle \pm i\left\vert
1\right\rangle )\right\} $ be the eigenstates of the Pauli matrices $\sigma
_{z}$, $\sigma _{x}$ and $\sigma _{y}$, respectively, where $s=\pm 1$ corresponds to
the eigenvalues. The target qubit state $\rho $ can be parametrized as
\begin{equation}
\rho =\left(
\begin{array}{cc}
1-a & k\sqrt{a(1-a)}e^{-i\Phi } \\
k\sqrt{a(1-a)}e^{i\Phi } & a%
\end{array}%
\right)     \label{rhorho}
\end{equation}%
where $a\in \left[ 0,1\right] $, $\Phi \in \left[ 0,2\pi \right] $ and $k\in %
\left[ 0,1\right] $. Since the Pauli-distance is invariant under the state
transformations $\rho (a,k,\Phi )\rightarrow \rho (1-a,k,\Phi )$ and $\rho
(a,k,\Phi )\rightarrow \rho (a,k,\Phi \pm n\pi /2)$ (with the integer $n$),
it is enough only to consider $a\in \left[ 0,1/2\right] $\ and $\Phi \in %
\left[ 0,\pi /2\right] $. In the Bloch representation, the density matrix is characterized by three positive numbers
\begin{eqnarray}
r_{ox} &=&2k\sqrt{a\left( 1-a\right) }\cos \Phi , \nonumber \\
r_{oy} &=&2k\sqrt{a\left( 1-a\right) }\sin \Phi , \nonumber \\
r_{oz} &=&1-2a.
\end{eqnarray}%
Thus we can present our results as follows.

\textbf{Theorem 4}.- Suppose $S:=\{\left\vert \varphi
_{1}\right\rangle =\left\vert \sigma _{\alpha }\right\rangle _{+},\left\vert
\varphi _{2}\right\rangle =\left\vert \sigma _{\alpha }\right\rangle
_{-},\left\vert \varphi _{3}\right\rangle =\left\vert \sigma _{\alpha
^{\prime }}\right\rangle _{+},\left\vert \varphi _{4}\right\rangle
=\left\vert \sigma _{\alpha ^{\prime }}\right\rangle _{-}\}$ with $\left\{
\alpha \neq \alpha ^{\prime }\neq \alpha ^{\prime \prime }|\alpha ,\alpha
^{\prime },\alpha ^{\prime \prime }=x,y,z\right\} $. If $r_{o\alpha
}r_{o\alpha ^{\prime }}\leq \mathfrak{m}$, the optimal Pauli distance is
given by%
\begin{equation}
D\left( \rho \right) =1-\frac{1}{\sqrt{2}}\left( 1+\sqrt{%
1-r_{o\alpha ^{\prime \prime }}^{2}}\right) ^{1/2}   \label{D4jie1}
\end{equation}%
and the corresponding optimal weights are given by
\begin{eqnarray}
p_{1} &=&\frac{1}{2}\left[ 1+\frac{r_{o\alpha }-r_{o\alpha ^{\prime }}}{%
\sqrt{1-r_{o\alpha ^{\prime \prime }}^{2}}}\right] -t,  \nonumber \\
p_{2} &=&\frac{1}{2}\left[ 1-\frac{r_{o\alpha }+r_{o\alpha ^{\prime }}}{%
\sqrt{1-r_{o\alpha ^{\prime \prime }}^{2}}}\right] -t,  \nonumber \\
p_{3} &=&\frac{r_{o\alpha ^{\prime }}}{\sqrt{1-r_{o\alpha ^{\prime \prime
}}^{2}}}+t,p_{4}=t,
\end{eqnarray}%
with%
\begin{equation}
t\subset \left[ 0,\frac{1}{2}-\frac{r_{o\alpha }+r_{o\alpha ^{\prime }}}{2%
\sqrt{1-r_{o\alpha ^{\prime \prime }}^{2}}}\right] .
\end{equation}%
If $r_{o\alpha }r_{o\alpha ^{\prime }}>\mathfrak{m}$, the optimal Pauli
distance is given by%
\begin{equation}
D\left( \rho \right) =1-\frac{1}{2}\left( 2+r_{o\alpha }+r_{o\alpha
^{\prime }}+\sqrt{4\mathfrak{m}+\left( r_{o\alpha }-r_{o\alpha ^{\prime
}}\right) ^{2}}\right) ^{1/2},    \label{D4jie2}
\end{equation}%
and the corresponding optimal weights are
\begin{eqnarray}
p_{1} &=&\frac{1}{2}\left[ 1+\frac{r_{o\alpha }-r_{o\alpha ^{\prime }}}{%
\sqrt{4\mathfrak{m}+\left( r_{o\alpha }-r_{o\alpha ^{\prime }}\right) ^{2}}}%
\right] , \nonumber \\
p_{3} &=&1-p_{1},p_{2}=p_{4}=0.  \label{P2zhi}
\end{eqnarray}

\textbf{Proof. }The squared fidelity between the target quantum state and
the set $S$ is given by%
\begin{equation}
F^{2}(\rho ,S)=\frac{1}{2}\left[ 1+r_{o\alpha }\left( p_{1}-p_{2}\right)
+r_{o\alpha ^{\prime }}\left( p_{3}-p_{4}\right) +\sqrt{\mathfrak{m}s}\right]
,
\end{equation}%
with%
\begin{equation}
s=2p_{1}p_{2}+2p_{3}p_{4}+p_{1}p_{3}+p_{1}p_{4}+p_{2}p_{3}+p_{2}p_{4}.
\end{equation}

\textit{Case 1}\textbf{:} $\mathfrak{m}=0$. In this case, the given target
quantum state $\rho $ is a pure state. Thus $\max F^{2}(\rho ,S)=\max \{%
\frac{1}{2}+\frac{r_{o\alpha }}{2},\frac{1}{2}+\frac{r_{o\alpha ^{\prime }}}{%
2}\}$, which corresponds to $p_{1}=1$ and $p_{3}=1$, respectively. This
result is a special expression of Eq. (\ref{P2zhi}).

\textit{Case 2}\textbf{:} $\mathfrak{m}\neq 0$. Here we establish the
Lagrangian function as%
\begin{equation}
L(p_{i},\lambda ,\lambda _{i})=1-F^{2}(\rho ,S)-\sum_{i=1}^{4}\lambda
_{i}p_{i}+\lambda (\sum_{i=1}^{4}p_{i}-1),
\end{equation}%
where $\lambda $ and $\lambda _{i}$ are the Lagrangian multiplier. The
Karush-Kuhn-Tucker conditions \cite{KKT} are given by%
\begin{equation}
\frac{\partial L}{\partial p_{i}}=0,\sum_{i=1}^{4}p_{i}=1,\lambda _{i}\geq
0,p_{i}\geq 0,\lambda _{i}p_{i}=0,
\end{equation}%
which can be further written as%
\begin{eqnarray}
-\frac{r_{o\alpha }}{2}-\frac{\sqrt{\mathfrak{m}}\left(
2p_{2}+p_{3}+p_{4}\right) }{2\sqrt{s}}+\lambda -\lambda _{1}=0,  \nonumber \\
\frac{r_{o\alpha }}{2}-\frac{\sqrt{\mathfrak{m}}\left(
2p_{1}+p_{3}+p_{4}\right) }{2\sqrt{s}}+\lambda -\lambda _{2}=0,  \nonumber \\
-\frac{r_{o\alpha ^{\prime }}}{2}-\frac{\sqrt{\mathfrak{m}}\left(
p_{1}+p_{2}+2p_{4}\right) }{2\sqrt{s}}+\lambda -\lambda _{3}=0,  \nonumber \\
\frac{r_{o\alpha ^{\prime }}}{2}-\frac{\sqrt{\mathfrak{m}}\left(
p_{1}+p_{2}+2p_{3}\right) }{2\sqrt{s}}+\lambda -\lambda _{4}=0,  \nonumber \\
\sum_{i=1}^{4}p_{i}=1,\lambda _{i}\geq 0,p_{i}\geq 0,\lambda
_{i}p_{i}=0,i=1,2,3,4.  \label{LL}
\end{eqnarray}%
Let us first consider the case $p_{i}\neq 0$ for $i=1,2,3,4$ which is
equivalent to $\lambda _{i}=0$. From $\frac{\partial L}{\partial p_{2}}-%
\frac{\partial L}{\partial p_{1}}=0$ and $\frac{\partial L}{\partial p_{4}}-%
\frac{\partial L}{\partial p_{3}}=0$, we can get%
\begin{eqnarray}
r_{o\alpha }\sqrt{s} &=&\sqrt{\mathfrak{m}}(p_{1}-p_{2}),  \nonumber \\
r_{o\alpha ^{\prime }}\sqrt{s} &=&\sqrt{\mathfrak{m}}(p_{3}-p_{4}).\label{L12}
\end{eqnarray}%
From $\frac{\partial L}{\partial p_{2}}-\frac{\partial L}{\partial p_{4}}=0$%
, we can get%
\begin{equation}
\left( r_{o\alpha }-r_{o\alpha ^{\prime }}\right) \sqrt{s}=\sqrt{\mathfrak{m}%
}(p_{1}-p_{2}-p_{3}+p_{4}),
\end{equation}%
which shows that Eq. (\ref{LL}) has countless solutions. For
convenience, we set $p_{4}=t$ with $t\in \left[ 0,1\right] $. Substituting $%
\sum_{i=1}^{4}p_{i}=1$ into Eq. (\ref{L12}) and Eliminating $\sqrt{s}$, it
follows that%
\begin{equation}
\left( r_{o\alpha }+r_{o\alpha ^{\prime }}\right) p_{1}+(r_{o\alpha
}-r_{o\alpha ^{\prime }})p_{2}=r_{o\alpha }\left( 1-2t\right) .
\end{equation}

\textit{Case 2.1}: $r_{o\alpha }\neq 0,r_{o\alpha }-r_{o\alpha
^{\prime }}\neq 0$. From the case, we have%
\begin{equation}
p_{2}=\frac{r_{o\alpha }\left( 1-2t\right) -\left( r_{o\alpha }+r_{o\alpha
^{\prime }}\right) p_{1}}{r_{o\alpha }-r_{o\alpha ^{\prime }}}.  \label{p21}
\end{equation}%
Substituting Eq. (\ref{p21}) into Eq. (\ref{L12}), we can obtain%
\begin{equation}
b_{1}p_{1}^{2}+b_{2}p_{1}+b_{3}=0,  \label{p122}
\end{equation}%
with%
\begin{eqnarray}
b_{1} &=&4\left( 2\mathfrak{m}+r_{o\alpha }^{2}+r_{o\alpha ^{\prime
}}^{2}\right) ,  \nonumber \\
b_{2} &=&-4\left( 1-2t\right) \left( 2\mathfrak{m}+r_{o\alpha
}^{2}+r_{o\alpha ^{\prime }}^{2}\right) , \nonumber \\
b_{3} &=&\left( 1-2t\right) ^{2}\left( 2\mathfrak{m}+r_{o\alpha
}^{2}+r_{o\alpha ^{\prime }}^{2}\right) -(r_{o\alpha }-r_{o\alpha ^{\prime
}})^{2}.
\end{eqnarray}%
Thus the roots of Eq. (\ref{p122}) read%
\begin{eqnarray}
p_{1} =\frac{-b_{2}\pm \sqrt{b_{2}^{2}-4b_{1}b_{3}}}{2b_{1}} 
=\frac{1}{2}\left[ 1\pm \frac{r_{o\alpha }-r_{o\alpha ^{\prime }}}{\sqrt{%
1-r_{o\alpha ^{\prime \prime }}^{2}}}\right] -t.  \label{p22}
\end{eqnarray}%
In order to uniquely determine the value of $p_{1}$, we substitute Eq. (\ref%
{p21}) and Eq. (\ref{p22}) into Eq. (\ref{L12}), and obtain the final $p_{1}$
value as%
\begin{equation}
p_{1}=\frac{1}{2}\left[ 1+\frac{r_{o\alpha }-r_{o\alpha ^{\prime }}}{\sqrt{%
1-r_{o\alpha ^{\prime \prime }}^{2}}}\right] -t.  \label{1P1}
\end{equation}%
Naturally, we can get%
\begin{eqnarray}
p_{2} &=&\frac{1}{2}\left[ 1-\frac{r_{o\alpha }+r_{o\alpha ^{\prime }}}{%
\sqrt{1-r_{o\alpha ^{\prime \prime }}^{2}}}\right] -t, \nonumber \\
p_{3} &=&\frac{r_{o\alpha ^{\prime }}}{\sqrt{1-r_{o\alpha ^{\prime \prime
}}^{2}}}+t,p_{4}=t.  \label{1P2}
\end{eqnarray}%
The necessary and sufficient condition of restriction $p_{i}\geqslant 0$
holds for $i=1,2,3,4$ is%
\begin{equation}
r_{o\alpha }r_{o\alpha ^{\prime }}\leq \mathfrak{m},t\subset \left[ 0,\frac{1%
}{2}-\frac{r_{o\alpha }+r_{o\alpha ^{\prime }}}{2\sqrt{1-r_{o\alpha ^{\prime
\prime }}^{2}}}\right] ,
\end{equation}%
Replace the optimal probability weights in Eq. (\ref{1P1}) and Eq. (\ref{1P2}%
) into Eq. (\ref{D1}), we have%
\begin{equation}
D_{S}\left( \rho \right) =1-\frac{1}{\sqrt{2}}\left( 1+\sqrt{%
1-r_{o\alpha ^{\prime \prime }}^{2}}\right) ^{1/2}.
\end{equation}%
If $r_{o\alpha }r_{o\alpha ^{\prime }}>\mathfrak{m}$, we can only find the
optimal distance from the following three cases (i) $p_{1}\neq 0$, $%
p_{2}\neq 0,p_{3}=p_{4}=0$; (ii)$\ p_{2}\neq 0,p_{3}\neq 0,p_{1}=p_{4}=0$;
(iii)$\ p_{1}\neq 0,p_{3}\neq 0,p_{2}=p_{4}=0$. In other words, these three
cases mean that only two numbers of $\{p_{i}\}$ are non-zero. For case (i),
the solution of Eq. (\ref{LL}) is%
\begin{eqnarray}
\tilde{p}_{1} &=&\frac{1}{2}\left[ 1+\frac{r_{o\alpha }}{\sqrt{2\mathfrak{m}%
+r_{o\alpha }^{2}}}\right] ,  \nonumber \\
\tilde{p}_{2} &=&1-\tilde{p}_{1},
\end{eqnarray}%
and%
\begin{equation}
\lambda _{3}=-\lambda _{4}=-\frac{r_{o\alpha ^{\prime }}}{2}<0,
\end{equation}%
without the restriction on $\lambda _{i}$. The solution contradicts the
condition $\lambda _{i}\geq 0$ for $i=1,2,3,4$. For case (ii), the solution
of Eq. (\ref{LL}) is%
\begin{eqnarray}
\tilde{p}_{2} &=&\frac{1}{2}\left[ 1-\frac{r_{o\alpha }+r_{o\alpha ^{\prime
}}}{\sqrt{4\mathfrak{m}+\left( r_{o\alpha }+r_{o\alpha ^{\prime }}\right)
^{2}}}\right] ,  \nonumber \\
\tilde{p}_{3} &=&1-\tilde{p}_{2},
\end{eqnarray}%
and%
\begin{equation}
\lambda _{1}=\lambda _{4}=\frac{1}{2}\left[ r_{o\alpha ^{\prime
}}-r_{o\alpha }-\sqrt{4\mathfrak{m}+\left( r_{o\alpha }+r_{o\alpha ^{\prime
}}\right) ^{2}}\right] <0,
\end{equation}%
without the restriction on $\lambda _{i}$. The solution also contradicts the
condition $\lambda _{i}\geq 0$ for $i=1,2,3,4$. For case (iii), the solution
of Eq. (\ref{LL}) is%
\begin{eqnarray}
p_{1} &=&\frac{1}{2}\left[ 1+\frac{r_{o\alpha }-r_{o\alpha ^{\prime }}}{%
\sqrt{4\mathfrak{m}+\left( r_{o\alpha }-r_{o\alpha ^{\prime }}\right) ^{2}}}%
\right] ,  \nonumber \\
p_{3} &=&1-p_{1},
\end{eqnarray}%
and%
\begin{equation}
\lambda _{2}=\lambda _{4}=\frac{1}{2}\left[ r_{o\alpha }+r_{o\alpha ^{\prime
}}-\sqrt{4\mathfrak{m}+\left( r_{o\alpha }-r_{o\alpha ^{\prime }}\right) ^{2}%
}\right] >0.
\end{equation}%
Obviously, only case (iii) has an analytical solution, and the optimal
distance is
\begin{equation}
D\left( \rho \right) =1-\frac{1}{2}\left( 2+r_{o\alpha }+r_{o\alpha
^{\prime }}+\sqrt{4\mathfrak{m}+\left( r_{o\alpha }-r_{o\alpha ^{\prime
}}\right) ^{2}}\right) ^{1/2}.
\end{equation}

\textit{Case 2.2}: $r_{o\alpha }=r_{o\alpha
^{\prime }}\neq 0$. From the case, we have%
\begin{equation}
p_{1}=\frac{1}{2}-t,p_{3}=\frac{1}{2}-p_{2},p_{4}=t.  \label{p1344}
\end{equation}%
Substituting Eq. (\ref{p1344}) into Eq. (\ref{L12}), we can obtain%
\begin{equation}
(r_{o\alpha }^{2}+m)(p_{2}+t-\frac{1}{2})^{2}+\frac{x^{2}}{2}=0.   \label{xp2}
\end{equation}%
By combining Eq. (\ref{L12}) and Eq. (\ref{xp2}), we can obtain that the unique solution of $p_{2}$ is
\begin{equation}
p_{2}=\frac{1}{2}-\frac{x}{\sqrt{1-r_{o\alpha ^{\prime\prime}}^{2}}}-t.  \label{xxp2}
\end{equation}%
Obviously, the analytic solution in Case 2.2 is a special expression of Eq. (\ref{1P1}) and  Eq. (\ref{1P2}).

\textit{Case 2.3}: $r_{o\alpha }= 0,r_{o\alpha }-r_{o\alpha
^{\prime }}\neq 0$. From the case, we have%
\begin{equation}
p_{1}=p_{2}=\frac{1-t-p_{3}}{2},p_{4}=t.  \label{p3123}
\end{equation}%
Substituting Eq. (\ref{p3123}) into Eq. (\ref{L12}), we can obtain%
\begin{equation}
(r_{o\alpha^{\prime} }^{2}+2m)(p_{3}-t)^{2}-r_{o\alpha^{\prime} }^{2}=0.   \label{xp3}
\end{equation}%
By combining Eq. (\ref{L12}) and Eq. (\ref{xp3}), we can obtain that the unique solution of $p_{3}$ is
\begin{equation}
p_{3}=\frac{r_{o\alpha^{\prime}}}{\sqrt{1-r_{o\alpha^{\prime\prime} }^{2}}}+t.  \label{xpp2}
\end{equation}%
So, the analytic solution in Case 2.3 is also a special case of Eq. (\ref{1P1}) and  Eq. (\ref{1P2}).
 The proof is completed.\hfill $\blacksquare $

According to theorem 4, we can give the best approximation of qubit states for the set composed of the eigenstates of any two Pauli matrices. Considering the easy preparation of these eigenstates, this theorem is of obvious experimental significance. In addition, we can find that our results are quite similar to those in \cite{CC3} based on trace norm. These two distance measures both illustrate the same fact that for Pauli set $S$, the optimal approximation of the target state can be achieved by up to three quantum states. However,  the optimal probability  with $B_{2}-$distances in \cite{CC3} is independent of $r_{o\alpha''}$, while ours with the fidelity obviously depends on it. This shows that for the optimal convex approximation,  different distance measures will lead to different understandings, which is the original motivation to study the problem with fidelity. 

As mentioned above, for the more general case of $N=4$, the similar method to theorem 2 is available, but the procedure is so tedious  that the final results would not be directly given. Next we would like to emphasize that  the case with $N> 4$ can be converted to the cases with $N\leq 4$. 

\textbf{Theorem 5}.- For $N> 4$, the optimal approximation is
determined by%
\begin{equation}
\min_{i_{1}<i_{2}<...<i_{N}}D(\rho ,\chi _{i_{1},i_{2},i_{3},i_{4}}\left(
\vec{p}\right) ),i_{1},i_{2},i_{3},i_{4}=1,2,3,4.
\end{equation}

\textbf{Proof}. Suppose we can select $N^{\prime }$ pure states from the set
$S$ to prepare the state denoted by
\begin{equation}
\rho _{o}^{\prime }=\sum\limits_{i=1}^{N^{\prime }}p_{i}^{\prime }\left\vert
\psi _{i}\right\rangle \left\langle \psi _{i}\right\vert
\end{equation}%
such that $D(\rho ,\rho _{o}^{\prime })$ achieves the optimal distance. In
Bloch representation, the above equation can be rewritten as
\begin{equation}
\mathbf{r_{o}^{\prime }}=\sum\limits_{i=1}^{N^{\prime }}p_{i}^{\prime }%
\mathbf{r_{i}}.
\end{equation}%
Caratheodory theorem \cite{Ca1,Ca2} states that if a point $x$ of $R^{d}$ lies in the
convex hull of a set $Q$, there is a subset $\tilde{Q}$ of $Q$ consisting of
$d+1$ or fewer points such that $x$ lies in the convex hull of $\tilde{Q}$.
An explicit version is that, if $\mathbf{r_{o}^{\prime }}$ is the convex
combination of $n$ Bloch vectors $\mathbf{r_{i}}$, there is a subset
consisting of not more than $4$ Bloch vectors $\mathbf{r_{j}}$ such that
\begin{equation}
\mathbf{r_{o}^{\prime }}=\sum\limits_{j=1}^{4}q_{i}\mathbf{r_{j}}.
\end{equation}%
with $\sum_{j=1}^{4}q_{j}=1$ and $q_{j}\geq 0$. Thus one can always find no
more than $4$ states from the selected $N^{\prime }$ pure states to exactly
prepare the state $\rho _{o}^{\prime }$ with some proper weights. This
implies that for $N>4$, the convex mixing of only $4$ states in $S$
is enough to achieve the optimal distance. In other words, the optimal state
$\rho _{o}$ can always be found from $\left\{ p_{i},\left\vert \psi
_{i}\right\rangle \right\} _{N}$ to make another pure-state decomposition
such that
\begin{equation}
\rho ^{\prime }=\sum_{j=1}^{4}q_{j}\left\vert \psi _{j}\right\rangle
\left\langle \psi _{j}\right\vert .
\end{equation}%
Therefore, we can directly consider all potential combinations of only $4$
pure states among the set $S$. The minimal distance will give our expected
optimal result.

\section{Examples}

In order to further prove the effectiveness of our theorems, we compare our
analytical results with the numerical results, where the target states $\rho $ and the
set $S$ are randomly generated. Similarly, the
target density  matrix $\rho $ is given in the form of Eq. (\ref{rhorho}).

(i) $N=2$. $S=\left\{ \left\vert \varphi _{1}\right\rangle ,\left\vert
\varphi _{2}\right\rangle \right\} $ with%
\begin{equation}
\left\vert \varphi _{1}\right\rangle =\left(
\begin{array}{c}
0.5143 \\
0.8317+0.2091i%
\end{array}%
\right) ,
\end{equation}%
and%
\begin{equation}
\left\vert \varphi _{2}\right\rangle =\left(
\begin{array}{c}
0.6950+0.5523i \\
0.3633+0.2827i%
\end{array}%
\right) .
\end{equation}%
In Fig. 1(a), the objective state $\rho $ is given by the randomly generated phase $\Phi
=0.4613\pi $. The optimal distance denoted by $%
D(\rho )$ versus $a\in \lbrack 0,1]$ is plotted in Fig. 1(a), where the dotted blue, solid red, dashed green, and dash-dot magenta lines correspond to the increasing order of variables $k=0.2$, $k=0.4$, $k=0.6$ and $k=0.8$, respectively. Similarly, the optimal distance $D(\rho )$ versus $k\in \lbrack 0,1]$ is plotted in Fig. 1(b), with the fixed parameter $a=0.8468$ and $\lbrace  \Phi=0,\pi/2,\pi,3\pi/2 \rbrace $. The optimal distance $D(\rho )$ versus $\Phi\in \lbrack 0,2\pi]$ is plotted in Fig. 1(c), with the fixed parameter $k=0.0131$ and $\lbrace a=0.2,0.4,0.6,0.8\rbrace$. These figures show the perfect consistency between the numerical and the analytical results, and
further support our theorem 1. 

\begin{figure}[tbp]
\centering
\subfigure[]{\includegraphics[width=0.32\columnwidth,height=1.4in]{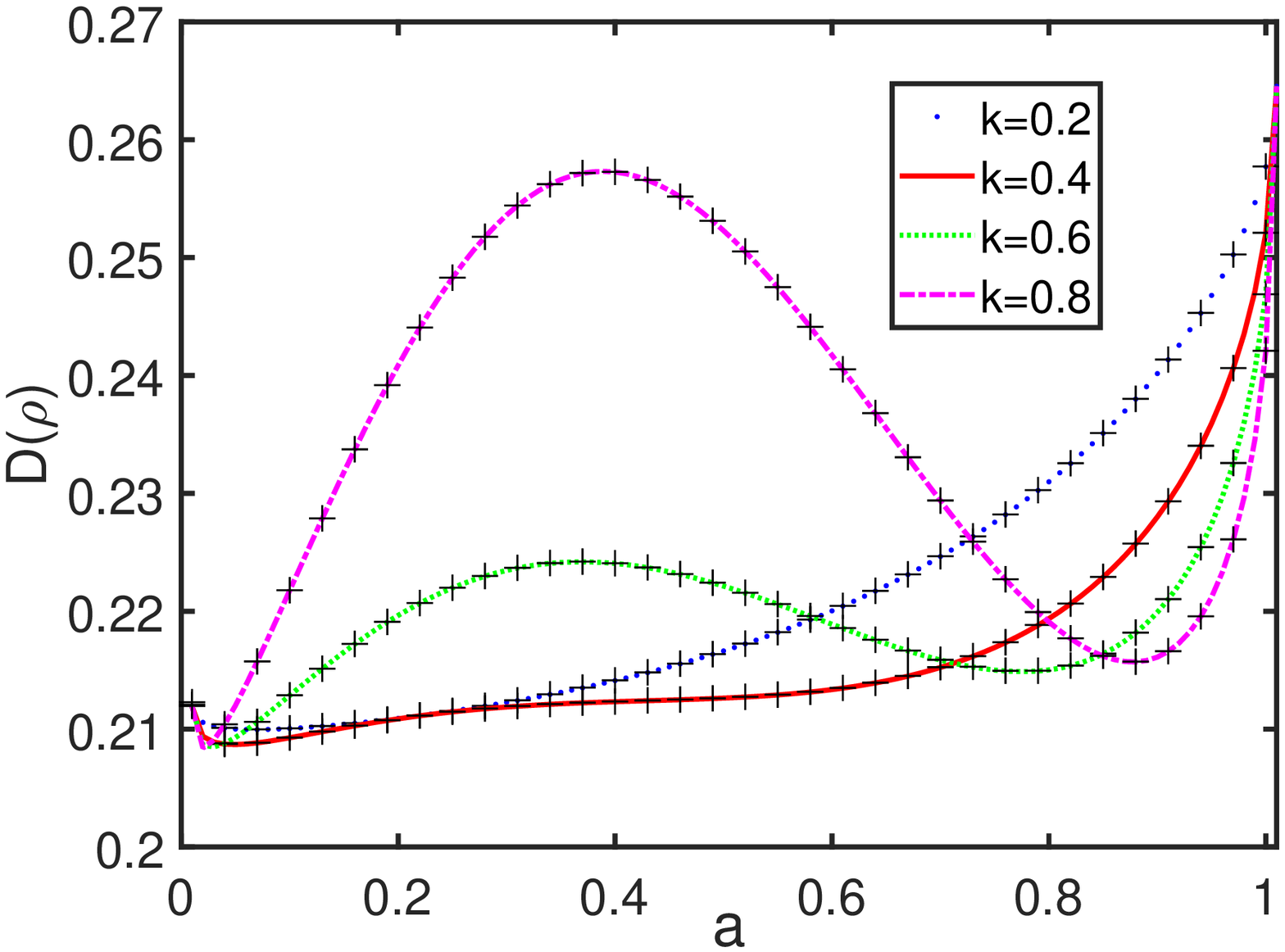}}
\subfigure[]{\includegraphics[width=0.32\columnwidth,height=1.4in]{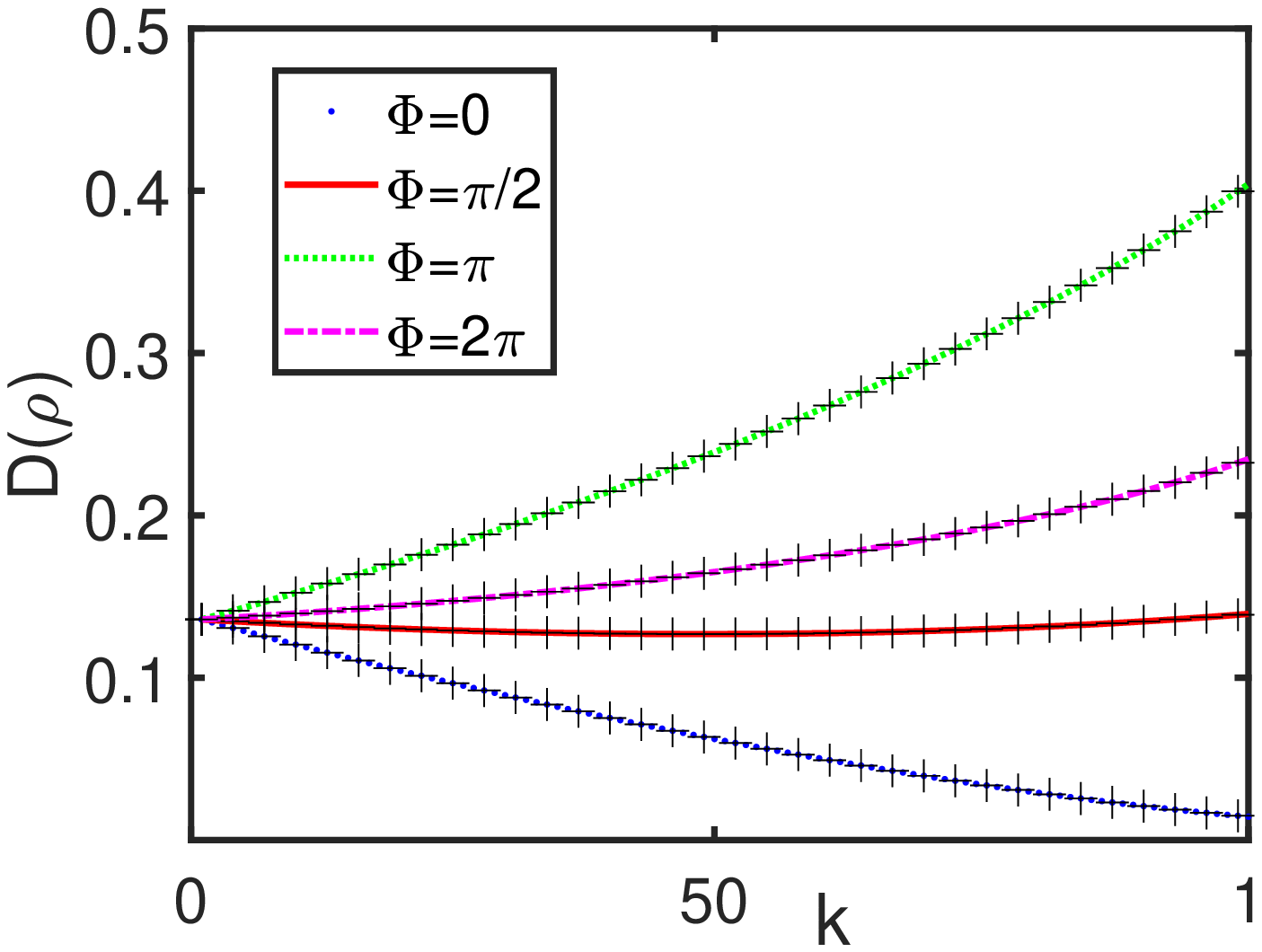}}
\subfigure[]{\includegraphics[width=0.32\columnwidth,height=1.4in]{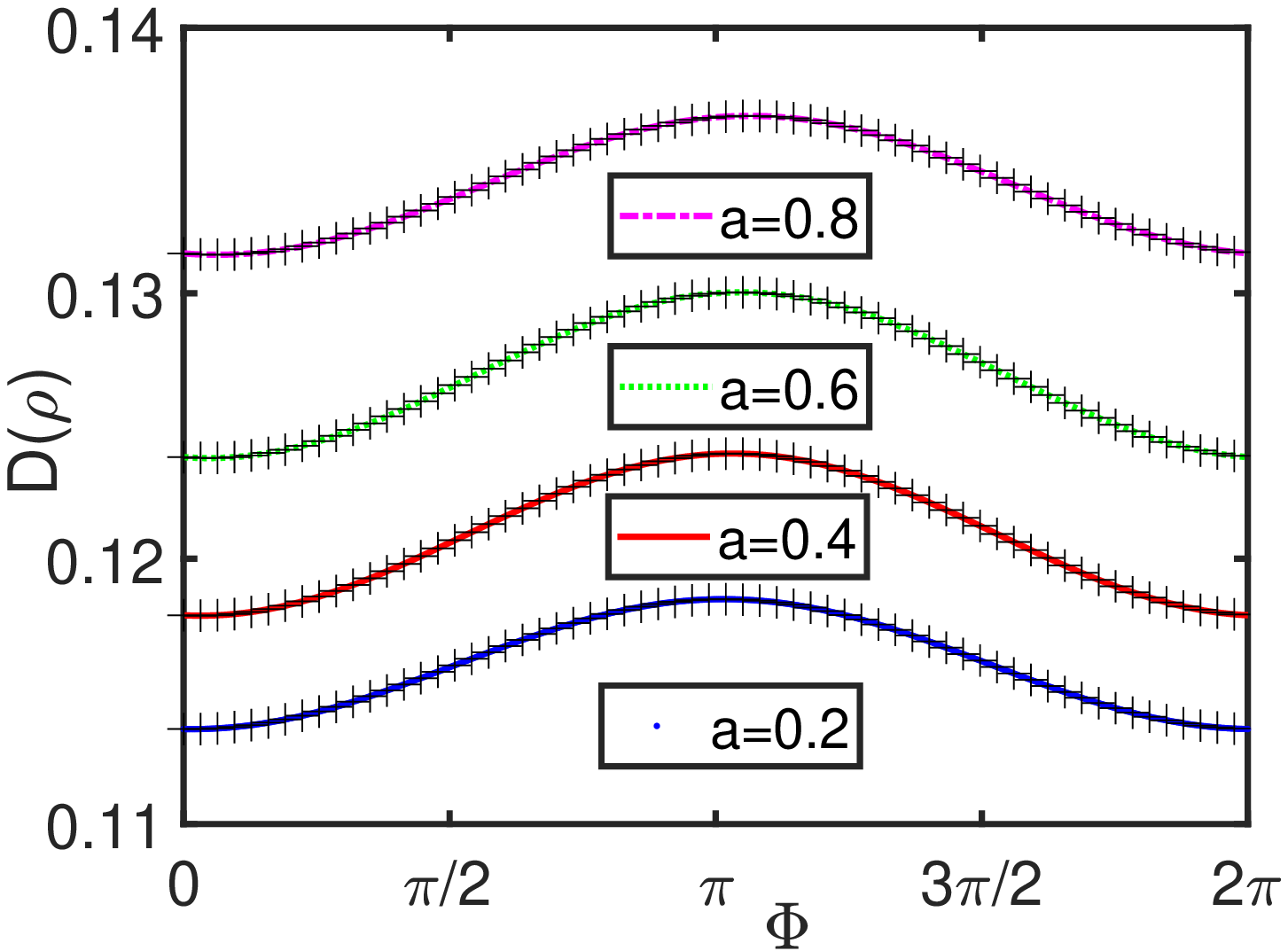}}
\caption{(color online) The optimal distance $D(\protect\rho )$ is plotted versus
various parameters $a$ in \textbf{(a)}, various parameters $k$ in \textbf{(b)} and various parameters $\Phi$ in \textbf{(c)}. The solid line corresponds to the strictly
analytical expressions of Eq. (\protect\ref{J1}), while the numerical
solutions are marked with "+". The numerical solution comes from the exhaustive method, that is, substituting $p_{1}=0:0.001:1$ and $p_{2}=1-p_{1}$ to calculate the value of each group and find the most suitable one.}
\end{figure}

(ii) $N=3$. $S=\left\{ \left\vert \varphi _{1}\right\rangle ,\left\vert
\varphi _{2}\right\rangle ,\left\vert \varphi _{3}\right\rangle \right\} $
with%
\begin{eqnarray}
\left\vert \varphi _{1}\right\rangle &=&\left(
\begin{array}{c}
0.5063+0.3025i \\
0.6829+0.4310i%
\end{array}%
\right) , \nonumber \\
\left\vert \varphi _{2}\right\rangle &=&\left(
\begin{array}{c}
0.1275+0.5888i \\
0.5452+0.5829i%
\end{array}%
\right) ,
\end{eqnarray}%
and%
\begin{equation}
\left\vert \varphi _{3}\right\rangle =\left(
\begin{array}{c}
0.0780+0.6594i \\
0.1059+0.7402i%
\end{array}%
\right) .
\end{equation}%
First we consider the objective state $\rho $ with randomly generated parameter $k=0.85$. The optimal distance denoted by $%
D(\rho )$ versus $a\in \lbrack 0,1]$ is plotted in Fig. 2(a), where the dotted blue, solid red, dashed green, and dash-dot magenta lines correspond to the increasing order of variables $\Phi=0$, $\Phi=\pi/2$, $\Phi=3\pi/2$ and $\Phi=2\pi$. Similarly, the optimal distance $D(\rho )$ versus $k\in \lbrack 0,1]$ is plotted in Fig. 2(b), with the fixed phase $\Phi=0.5318\pi $ and $\lbrace a=0.2,0.4,0.6,0.8 \rbrace$. The optimal distance $D(\rho )$ versus $\Phi\in \lbrack 0,2\pi]$ is plotted in Fig. 1(c), with the fixed parameter $a=0.63$ and $\lbrace k=0.2,0.4,0.6,0.8 \rbrace$. These figures validate our theorem
2 based on the perfect consistency.
\begin{figure}[tbp]
\centering
\subfigure[]{\includegraphics[width=0.32\columnwidth,height=1.4in]{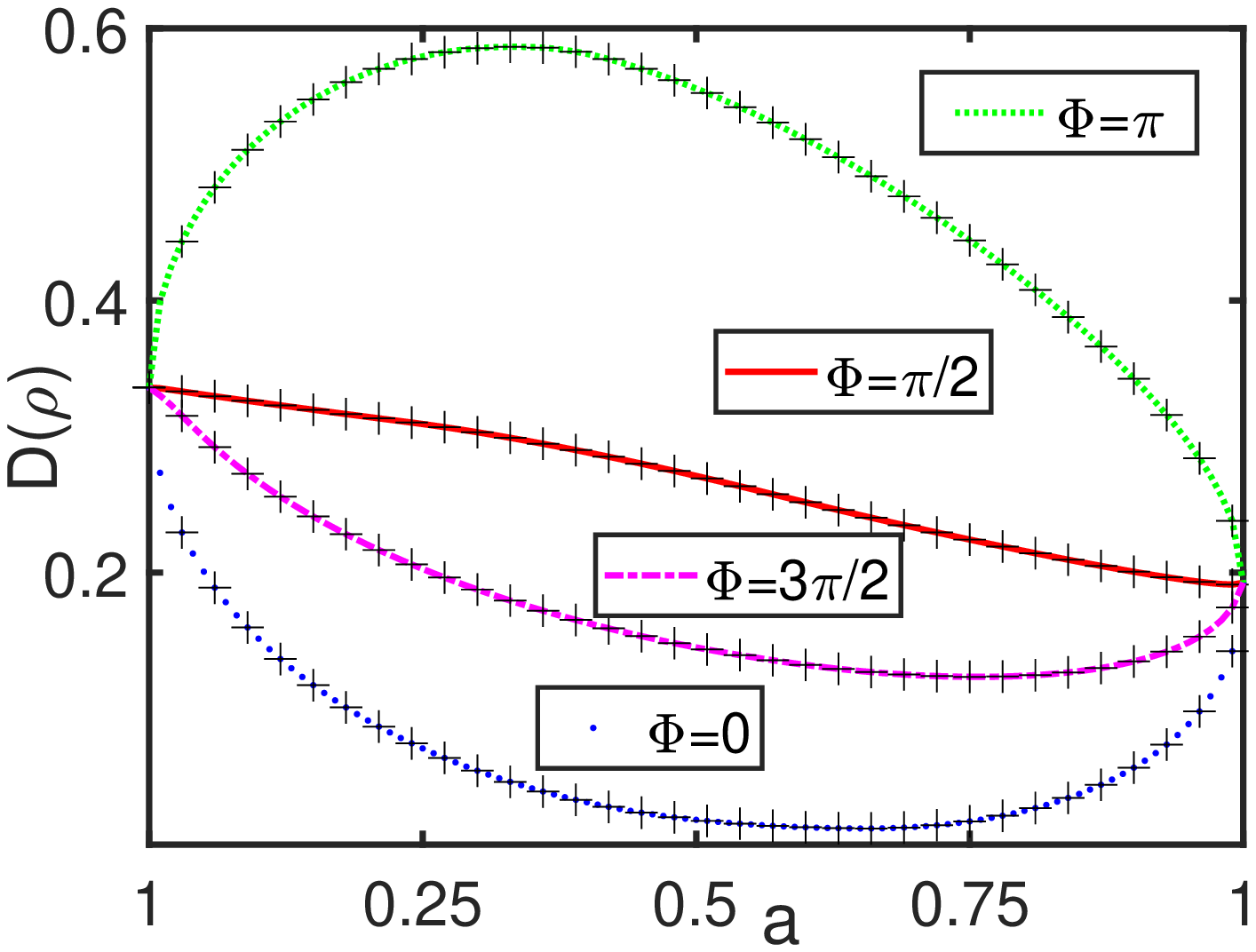}}
\subfigure[]{\includegraphics[width=0.32\columnwidth,height=1.4in]{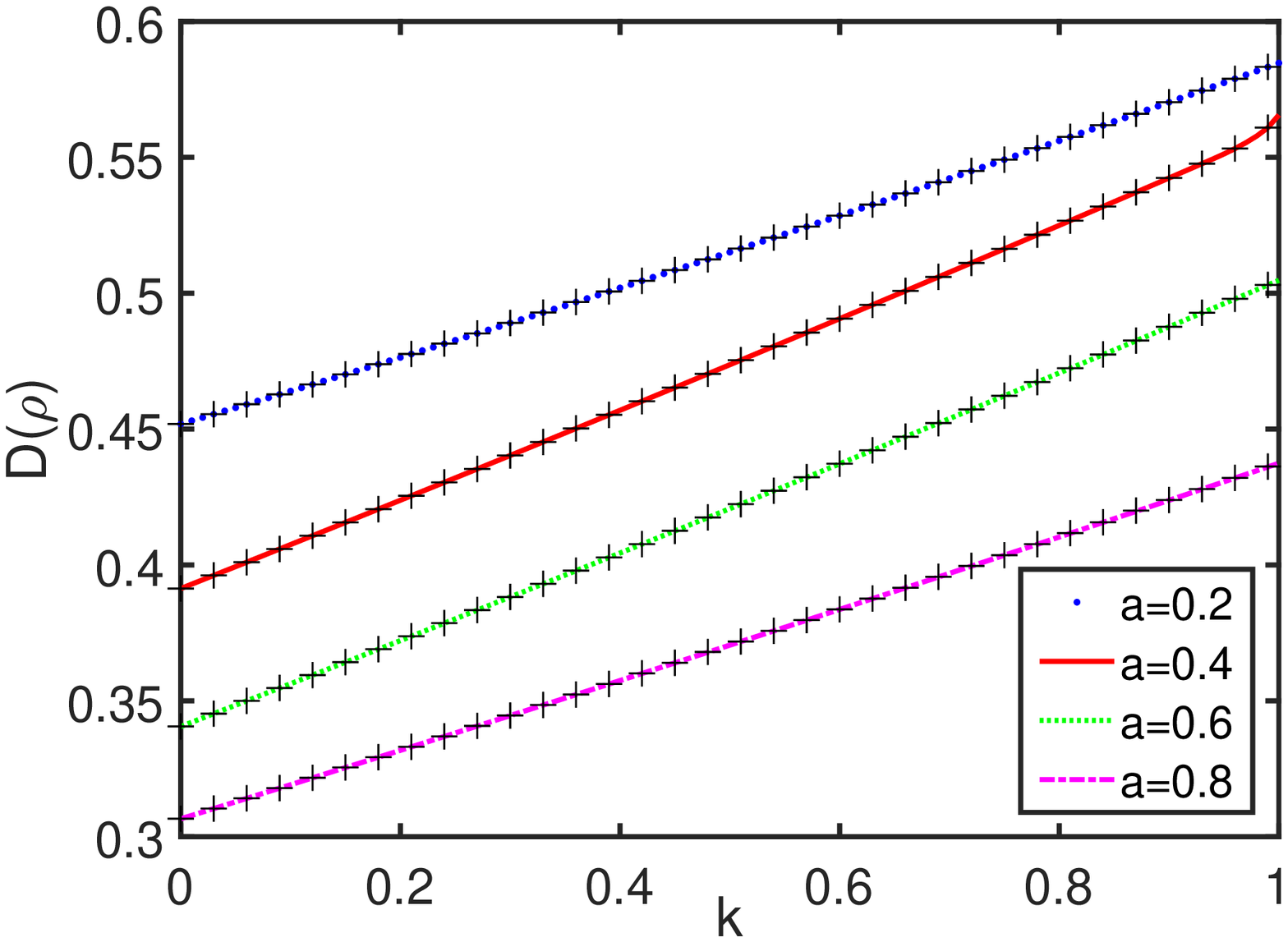}}
\subfigure[]{\includegraphics[width=0.32\columnwidth,height=1.4in]{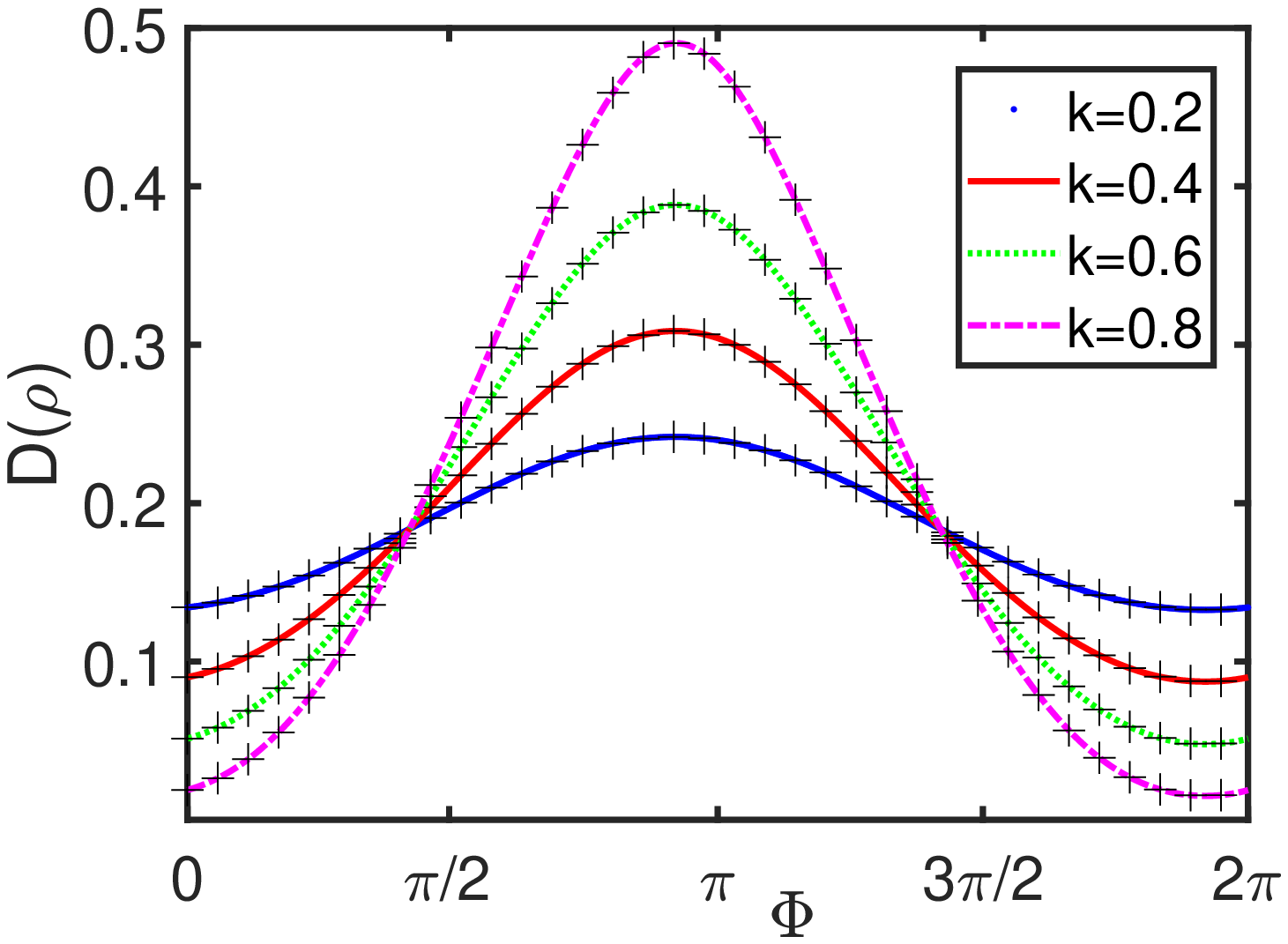}}
\caption{(color online) The optimal distance $D(\protect\rho )$ versus
various parameters $a$ in \textbf{(a)}, various parameters $k$ in \textbf{(b)} and various parameters $\Phi$ in \textbf{(c)}. The solid line corresponds to the strictly
analytical expressions of Eq. (\protect\ref{J22a1}) and Eq. (%
\protect\ref{J22a}), while the numerical solutions are marked with "+". The numerical solution comes from the exhaustive method, that is, substituting $p_{1}=0:0.001:1$, $p_{2}=0:0.001:p_{1}$ and $p_{3}=1-p_{1}-p_{2}$ to calculate the value of each group and find the most suitable one.}
\end{figure}

(iii) $N=4$. $S=\{\left\vert \sigma _{x }\right\rangle _{+},\left\vert \sigma _{x }\right\rangle
_{-},\left\vert \sigma _{z}\right\rangle _{+},\left\vert \sigma _{z}\right\rangle _{-}\}$. In Fig. 3(a), the objective state $\rho $ is given by the randomly generated $k=0.5910$. The optimal distance denoted by $%
D(\rho )$ versus $a\in \lbrack 0,1]$ is plotted in Fig. 3(a), where the dotted blue, solid red, dashed green, and dash-dot magenta lines correspond to the increasing order of variable $\Phi=0$, $\Phi=\pi/4$, $\Phi=\pi/3$ and $\Phi=\pi/2$, respectively. Similarly, the optimal distance $D(\rho )$ versus $k\in \lbrack 0,1]$ is plotted in Fig. 3(b), with the fixed phase $\Phi=0.4047\pi$ and $\lbrace  a=0.2,0.4,0.6,0.8\rbrace $. The optimal distance $D(\rho )$ versus $\Phi\in \lbrack 0,\pi/2]$ is plotted in Fig. 1c, with the fixed parameter $a=0.1145$ and $\lbrace k=0.2,0.4,0.6,0.8\rbrace$. These figures support our theorem
4.
\begin{figure}[tbp]
\centering
\subfigure[]{\includegraphics[width=0.32\columnwidth,height=1.4in]{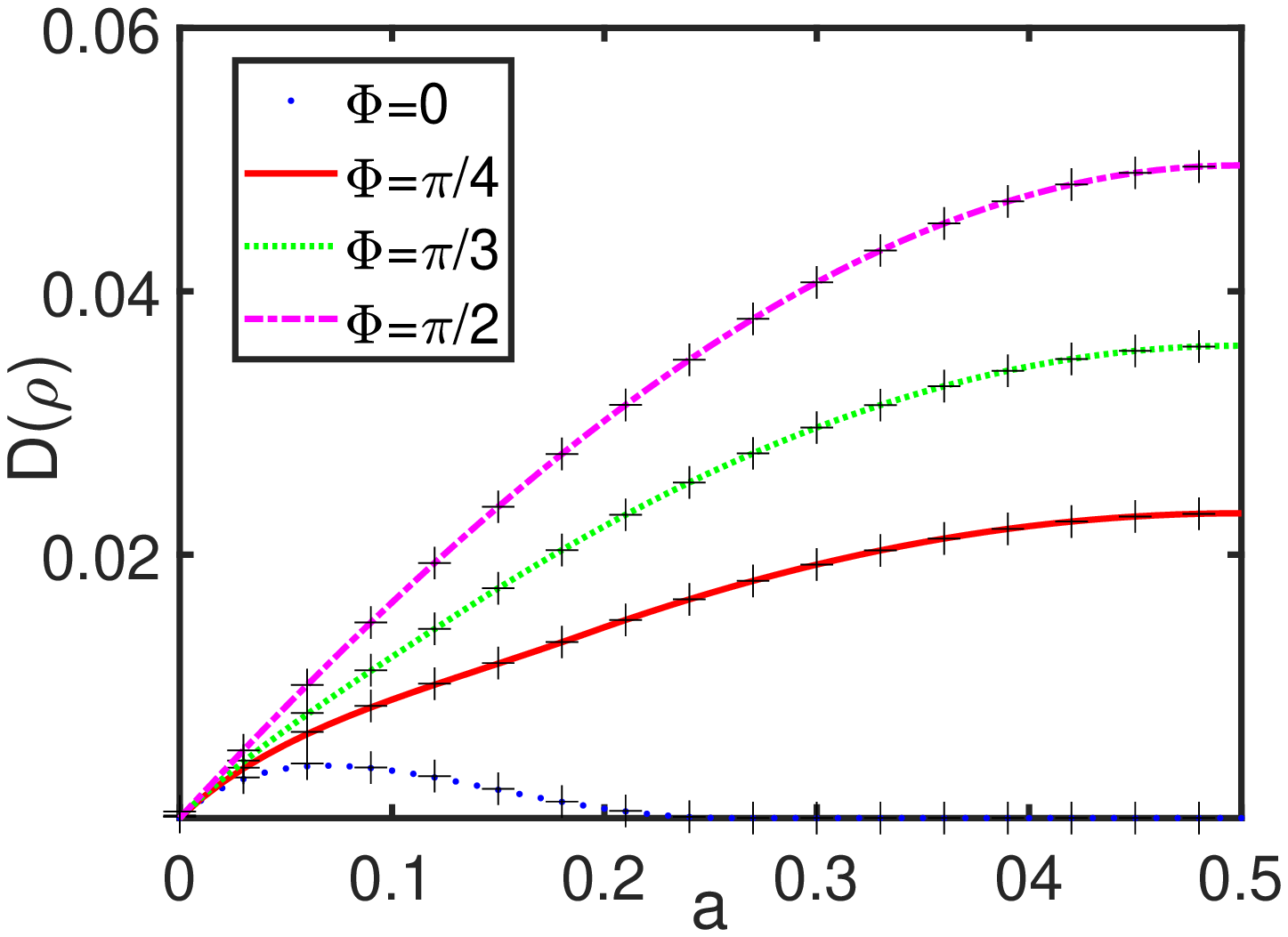}}
\subfigure[]{\includegraphics[width=0.32\columnwidth,height=1.4in]{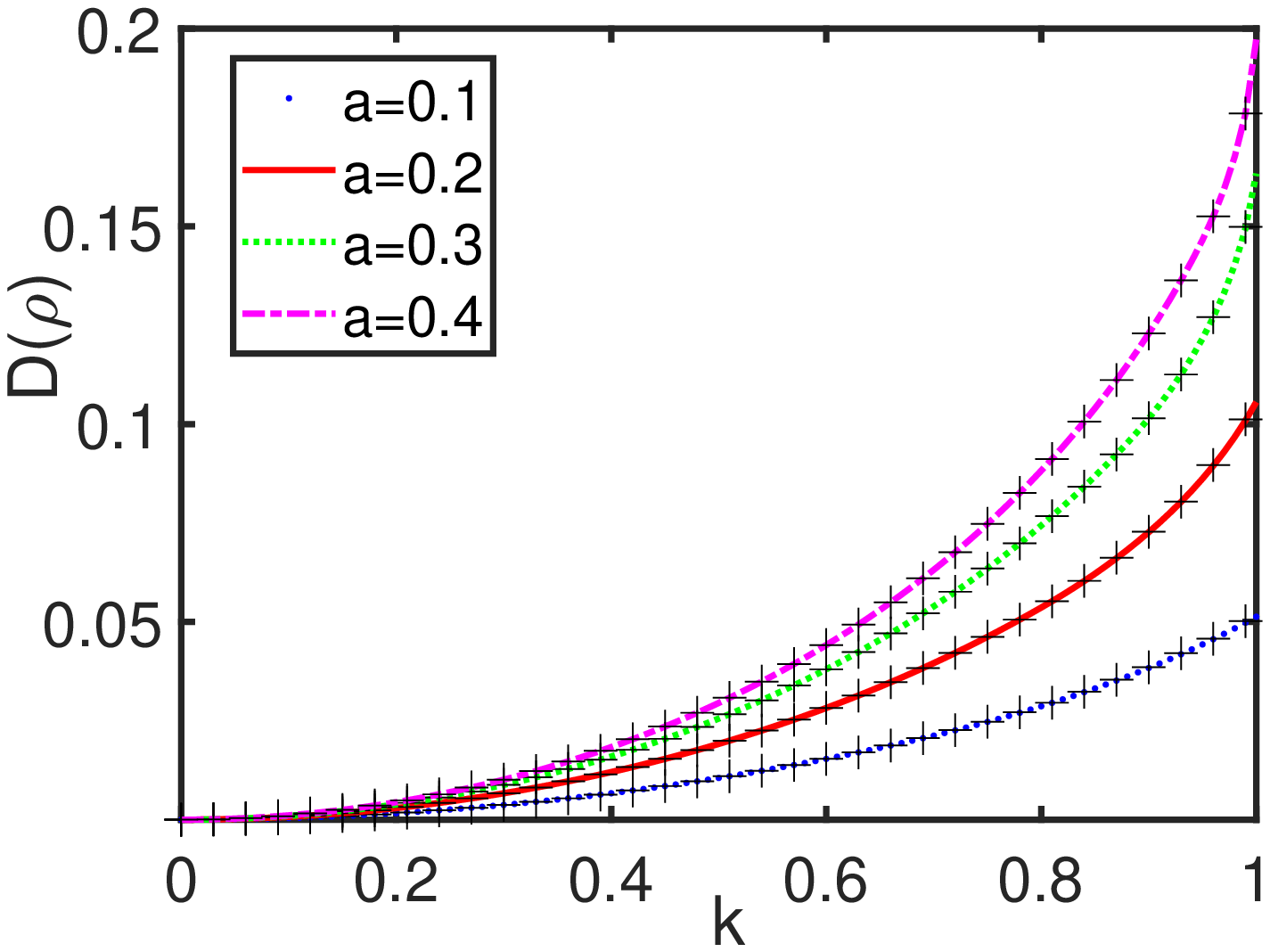}}
\subfigure[]{\includegraphics[width=0.32\columnwidth,height=1.4in]{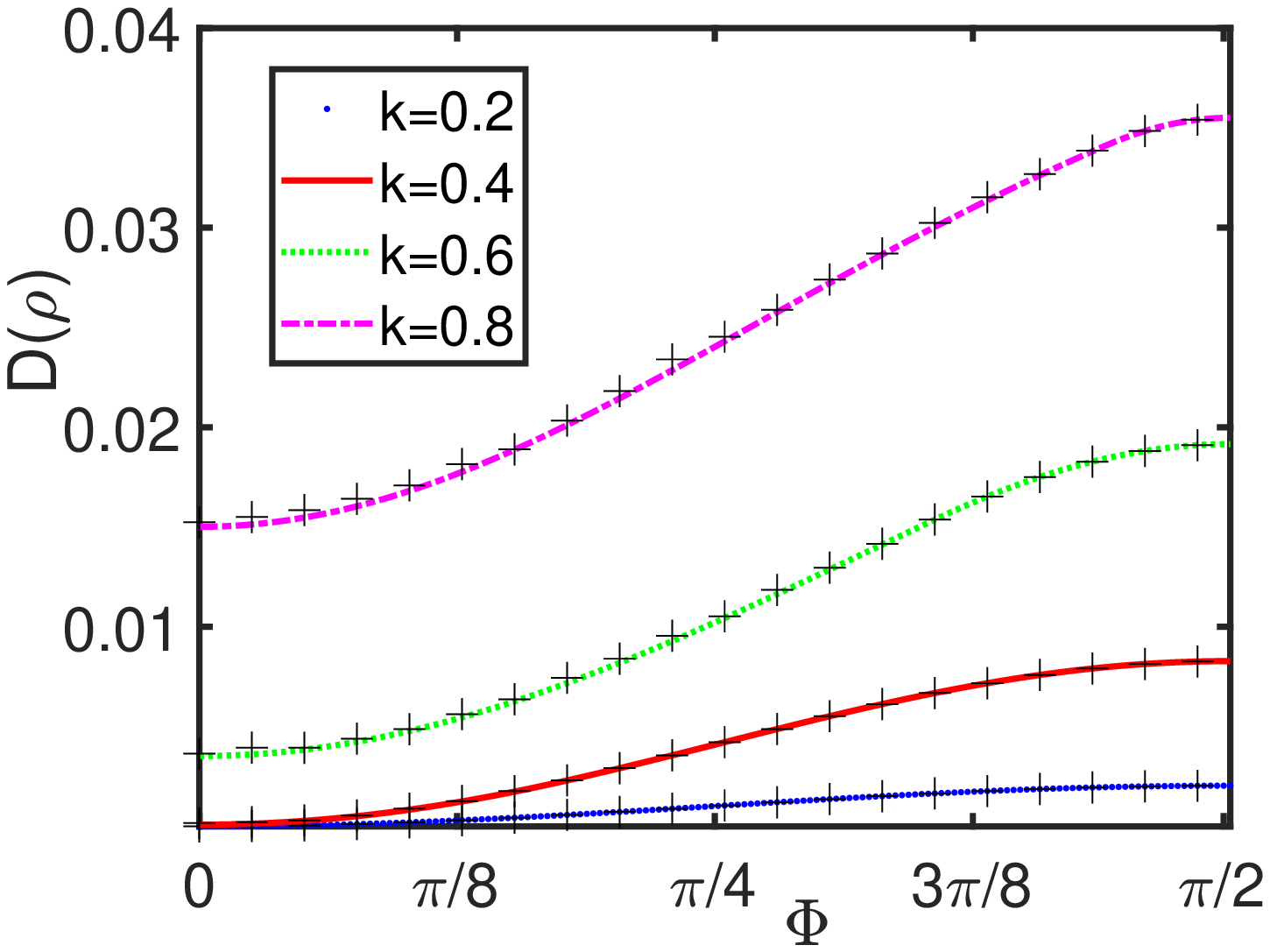}}
\caption{(color online) The optimal distance $D(\protect\rho )$  versus
various parameters $a$ in \textbf{(a)}, various parameters $k$ in \textbf{(b)} and various parameters $\Phi$ in \textbf{(c)}. The solid line corresponds to the strictly
analytical expressions of Eq. (\protect\ref{D4jie1}) and Eq. (\protect\ref{D4jie2}), while the numerical
solutions are marked with "+". The numerical solution comes from the exhaustive method, that is, substituting $p_{1}=0:0.001:1$, $p_{2}=0:0.001:1-p_{1}$, $p_{3}=0:0.001:1-p_{1}-p_{2}$ and $p_{4}=1-p_{1}-p_{2}-p_{3}$ to calculate the value of each group and find the most suitable one. }
\end{figure}

\section{ Discussion and conclusions}

Before the end, we want to emphasize that we mainly give a method to calculate how to use the available set of qubit pure states to optimally approximate a target qubit state. This method  can be directly used for mixed-state set since all qubit states can be given in the Bloch representation.  For example, if condition $\mathbf{r}_{i}^{T}\mathbf{r}_{i}=1$ in Eq. (\protect\ref{fsq}) is changed to condition $\mathbf{r}_{i}^{T}\mathbf{r}_{i}\leqslant1$, we can generalize the result from pure qubit state to mixed qubit state. However, one can find that the result becomes very tedious, so we only present the results for pure states. It is a challenge to use the fidelity for the problem in high dimensional cases. The reason is that the fidelity for high dimensional quantum states is hard to calculate in analytical form. This directly prevents the generalization to high dimensional systems.

To summarize, we have solved the best approximation of a qubit state with some
given pure states, that is, optimally approximating a given qubit density
matrix by the convex summation of pure states with $N=2$, $N=3$ and the
special case of $N=4$. For the general case of $N=4$, we can use the same
method as theorem 2. The optimal approximation with $N> 4$ pure states
can be converted into the problem with not more than four pure states. The
numerical comparisons illustrate the validity of our analytic results.

\ack

This work was supported by the National Natural Science Foundation of China
under Grant No.11775040, No. 12011530014 and No.11375036, and the
Fundamental Research Fund for the Central Universities under Grant No.
DUT20LAB203.

\end{document}